\documentclass{article}
\usepackage[T1]{fontenc} 
\usepackage[utf8]{inputenc} 
\usepackage{ismir,amsmath,cite,url}
\usepackage{graphicx}
\usepackage{color}
\usepackage{ amssymb }
\usepackage{booktabs}
\usepackage{lineno}

\title{AccoMontage: Accompaniment Arrangement via Phrase Selection and Style Transfer}

\twoauthors
  {Jingwei Zhao} {Music X Lab, NYU Shanghai \\  \href{mailto: jz4807@nyu.edu}{\tt jz4807@nyu.edu}}
  {Gus Xia} {Music X Lab, NYU Shanghai \\ \href{mailto: gxia@nyu.edu}{\tt gxia@nyu.edu}}

\def\authorname{J. Zhao, and G. Xia}

\usepackage[bookmarks=false,pdfauthor={\authorname},pdfsubject={\papersubject},hidelinks]{hyperref}

\usepackage{bbm}

\begin{document}

\maketitle
\begin{abstract}
Accompaniment arrangement is a difficult music generation task involving intertwined constraints of melody, harmony, texture, and music structure. Existing models are not yet able to capture all these constraints effectively, especially for long-term music generation. To address this problem, we propose {\itshape AccoMontage}, an accompaniment arrangement system for \textit{whole pieces} of music through unifying phrase selection and neural style transfer.\footnote{Codes and demos at \href{https://github.com/zhaojw1998/AccoMontage}{https://github.com/zhaojw1998/AccoMontage}.} We focus on generating piano accompaniments for folk/pop songs based on a lead sheet (i.e., melody with chord progression). Specifically, AccoMontage first retrieves phrase montages from a database while recombining them structurally using dynamic programming. Second, chords of the retrieved phrases are manipulated to match the lead sheet via style transfer. Lastly, the system offers controls over the generation process. In contrast to pure learning-based approaches, AccoMontage introduces a novel hybrid pathway, in which rule-based optimization and deep learning are both leveraged to complement each other for high-quality generation. Experiments show that our model generates well-structured accompaniment with delicate texture, significantly outperforming the baselines.
\end{abstract}
\section{Introduction}\label{sec:introduction}
\textit{Accompaniment arrangement} refers to the task of reconceptualizing a piece by composing the accompaniment part given a lead sheet (a lead melody with a chord progression). When designing the texture and voicing of the accompaniment, arrangers are simultaneously dealing with the \textit{constraints} from the original melody, chord progression, and other structural information. This constrained composition process is often modeled as a \textit{conditioned generation problem} in music automation. 

Despite recent promising advances in deep music generative models \cite{yang2019deep, 9031528, huang2018music, dhariwal2020jukebox, ren2020popmag, zhu2018xiaoice, yang2017midinet}, existing methods cannot yet generate musical accompaniment while capturing the aforementioned constraints effectively. Specifically, most algorithms fall short in preserving the fine granularity and structure of accompaniment in the long run. Also, it is difficult to explicitly control the generation process. We argue that these limits are mainly due to the current \textit{generation from scratch} approach.  In composition practice, however, arrangers often resort to existing pieces as accompaniment references. For example, a piano accompanist can improvise through off-the-shelf textures while transferring them into proper chords, which is essentially re-harmonizing a reference piece to fit a query lead sheet. In this way, the coherence and structure of the accompaniment are preserved from the reference pieces, and musicians also have control over what reference to choose.

To this end, we contribute {\itshape AccoMontage}, a \textit{generalized template-based} approach to 1) given a lead sheet as the query, search for proper accompaniment phrases as the reference; 2) re-harmonize the selected reference via style transfer to accompany the query. We model the search stage as an optimization problem on the graph, where nodes represent candidate phrases in the dataset and edges represent inter-phrase transitions. Node scores are defined in a rule-based manner to reveal query-reference fitness, while edge scores are learned by contrastive learning to reveal smoothness of phrase transitions. As for the re-harmonization stage, we adopt the chord-texture disentanglement and transfer method in \cite{wang2020learning, wang2020pianotree}. 

The current system focuses on arranging piano accompaniments for a full-length folk or pop song. Experimental results show that the generated accompaniments not only harmonize well with the melody but also contain more intra-phrase coherence and inter-phrase structure compared to the baselines. In brief, our contributions are:
\begin{itemize}

\item \textbf{A generalized template-based approach}: A novel hybrid approach to generative models, where searching and deep learning are both leveraged to complement each other and enhance the overall generation quality. This strategy is also useful in other domains.
   
\item \textbf{The AccoMontage system}: A system capable of generating long-term and structured accompaniments for full-length songs. The arranged accompaniments have state-of-the-art quality and are significantly better than existing pure learning-based and template-based baselines. 
   
\item \textbf{Controllable music generation}: Users can control the generation process by pre-filtering of two texture features: rhythm density and voice number.
\end{itemize}

\section{Related Work}
We review three topics related to symbolic accompaniment arrangement: conditional music generation, template-based arrangement, and music style transfer.

\subsection{Conditional Music Generation}
Conditional music generation takes various forms, such as generating chords conditioned on the melody \cite{simon2008mysong, lim2017chord}, generating melody on the underlying chords \cite{zhu2018xiaoice, yang2017midinet}, and generating melody from metadata and descriptions \cite{zhang2020butter}. In particular, accompaniment arrangement refers to generating accompaniment conditioned on the lead sheet, and this topic has recently drawn much research attention. We even see tailored datasets for piano arrangement tasks \cite{wang2020pop909}. 

For accompaniment arrangement, existing models that show satisfied arrangement quality typically apply only to \textit{short} clips. GAN and VAE-based models are used to maintain inter-track music dependency \cite{dong2018musegan, liu2018lead, jia2019impromptu}, but limit music generation within 4 to 8 bars. Another popular approach is to generate longer accompaniment in a seq2seq manner with attention \cite{wang2020learning, ren2020popmag, zhu2018xiaoice}, but can easily converge to repetitive textural patterns in the long run. On the other hand, models that arrange for complete songs typically rely on a library of fixed elementary textures and often fail to generalize \cite{chen2013automatic, wu2016emotion, liu2012polyphonic}. This paper aims to unite both high-quality and long-term accompaniment generation in one system, where ``long-term'' refers to full songs (32 bars and more) with dependencies to intra-phrase melody and chord progression, and inter-phrase structure.

\subsection{Template-based Accompaniment Arrangement}
The use of existing compositions to generate music is not an entirely new idea. Existing template-based algorithms include learning-based unit selection \cite{bretan2016unit,xia_2018}, rule-based matching \cite{chen2013automatic, wu2016emotion}, and genetic algorithms \cite{liu2012polyphonic}. For accompaniment arrangement, a common problem lies in the difficulty to find a perfectly matched reference especially when the templates contain rich textures with non-chordal tones. Some works choose to only use basic accompaniment patterns to avoid this issue \cite{chen2013automatic, wu2016emotion, liu2012polyphonic}. In contrast, our study addresses this problem by applying the style transfer technique on a selected template to improve the fitness between the accompaniment and the lead sheet. We name our approach after \textit{generalized} template matching.

\subsection{Music Style transfer}
Music style transfer \cite{dai2018music} is becoming a popular approach for controllable music generation. Through music-representation disentanglement and manipulation, users can transfer various factors of a reference music piece, including pitch contour, rhythm pattern, chord progression, polyphonic texture, etc \cite{wang2020learning,yang2019deep}. Our approach can be seen as an extension of music style transfer in which the ``reference search'' step is also automated.

\section{Methodology}
The AccoMontage system uses a generalized template-based approach for piano accompaniment arrangement. The input to the system is a lead sheet of a complete folk/pop song with phrase labels, which we call a {\itshape query}. The search space of the system is a MIDI dataset of piano-arranged pop songs. In general, we can derive the chord progression and phrase labels of each song in the dataset by MIR algorithms. In our case, the chords are extracted by \cite{wang2020pop909} and the phrases are labeled manually \cite{dai2020automatic}. We refer to each phrase of associated accompaniment, melody, and chords as a {\itshape reference}. For the rest of this section, we first introduce the feature representation of the AccoMontage system in Section \ref{sec-feature}, and then describe the main pipeline algorithms in Section \ref{sec-phrase} and \ref{styletransfer}. Finally, we show how to further control the arrangement process in Section \ref{pre-filtering-control}.

\subsection{Feature Representation}\label{sec-feature}
Given a lead sheet as the query, we represent it as a sequence of ternary tuples:
\begin{equation}
    q = \{\left(q_i^{\mathrm{mel}}, q_i^{\mathrm{chord}}, q_i^{\mathrm{label}}\right)\}_{i=1}^{n},
\end{equation}
where $q_i^{\mathrm{mel}}$, the melody feature of query phrase $i$, is a sequence of 130-D one-hot vectors with 128 MIDI pitches plus a hold and a rest state \cite{roberts2018hierarchical}; $q_i^{\mathrm{chord}}$, the chord feature aligned with $q_i^{\mathrm{mel}}$, is a sequence of 12-D chromagram vectors \cite{yang2019deep,9031528}; $q_i^{\mathrm{label}}$ is a phrase label string denoting within-song repetition and length in bar, such as $\mathtt{A8}$, $\mathtt{B8}$, etc. \cite{dai2020automatic}. $n$ is the number of phrases in lead sheet $q$.

We represent the accompaniment reference space as a collection of tuples: 
\begin{equation}
    r = \{\left(r_i^{\mathrm{mel}}, r_i^{\mathrm{chord}}, r_i^{\mathrm{acc}}\right)\}_{i=1}^N,
\end{equation}
where $r_i^{\mathrm{mel}}$, and $r_i^{\mathrm{chord}}$ are the melody and the chord feature of the $i$-th reference phrase, represented in the same format as in the query phrases; $r_i^{\mathrm{acc}}$ is the accompaniment feature, which is a 128-D piano-roll representation the same as \cite{wang2020learning}. $N$ is the volume of the reference space.

\subsection{Phrase Selection}\label{sec-phrase}
Assuming there are $n$ phrases in the query lead sheet, we aim to find a reference sequence:
\begin{equation}
    {\bf x} = [x_1, x_2, \cdots, x_n],
\end{equation}
where we match reference $x_i$ to the $i$-th phrase $q_i$ in our query; $x_i \in r$ and has the same length as $q_i$.

Given the query's phrase structure, the reference space forms a graph of layered structures shown as \figref{selection}. Each layer consists of equal-length reference phrases and consecutive layers are fully connected to each other. Each node in graph describes the fitness between $x_i$ and $q_i$, and each edge evaluates the transition from $x_{i}$ to $x_{i+1}$. A complete selection of reference phrases corresponds to a path that traverses through all layers. To evaluate a path, We design a {\itshape fitness model} and a {\itshape transition model} as follows.

\subsubsection{Phrase Fitness Model}\label{fitmodel}
We rely on the phrase fitness model to evaluate if a reference accompaniment phrase matches a query phrase. Formally, we define the fitness model $f(x_i, q_i)$ as follows:
\begin{equation}\label{sim}
\begin{aligned}
    f(x_i, q_i) &= \alpha \mathrm{sim}(x_i^{\mathrm{rhy}}, q_i^{\mathrm{rhy}}) \\
                &+ \beta \mathrm{sim}(\mathrm{T}(x_i^{\mathrm{chord}}), \mathrm{T}(q_i^{\mathrm{chord}})),
\end{aligned}
\end{equation}
where $\mathrm{sim}(\cdot, \cdot)$ measures the similarity between two inputs. In our work, we use the cosine similarity. $\mathrm{T}(\cdot)$ is the Tonal Interval Vector (TIV) operator that maps a chromagram to a 12-D tonal interval space whose geometric properties concur with harmonic relationships of the tonal system \cite{bernardes2016multi}. $x_i^{\mathrm{rhy}}$ and $q_i^{\mathrm{rhy}}$ are both rhythm features, which condense the original 130-D melody feature to 3-D that denotes an onset of any pitch, a hold state, and rest \cite{yang2019deep}. $x_i^{\mathrm{chord}}$ and $q_i^{\mathrm{chord}}$ are chord features (chromagram) defined in Section \ref{sec-feature} and we further augment the reference space by transposing phrases to all 12 keys. While computing the similarity, we consider the rhythm feature and TIV as 2-D matrices each with channel number 3 and 12. We calculate the cosine similarity of both features by feeding in their channel-flattened vectors. 

Note that in Eq \eqref{sim}, we compare only the rhythm and chord features for query-reference matching. The underlying assumption is that \textit{if lead sheet $A$ is similar to another lead sheet $B$ in rhythm and chord progression, then $B$'s accompaniment will be very likely to fit $A$ as well}. 

\subsubsection{Transition Model}\label{transcore}
We exploit the transition model to reveal the inter-phrase transition and structural constraints.  Formally, we define the transition score between two reference accompaniment phrases $t(x_i, x_{i+1})$ as follows:
\begin{equation}\label{tran}
\begin{aligned}
     t(x_i, x_{i+1}) &= \mathrm{sim}(W_1 x_i^{\mathrm{txt}}, W_2 x_{i+1}^{\mathrm{txt}}) \\ &+ \mathrm{form}(x_i, x_{i+1}).
\end{aligned}
\end{equation}

The first term in Eq \eqref{tran} aims to reveal the transition naturalness of the polyphonic \textit{texture} between two adjacent phrases. Instead of using rule-based heuristics to process texture information, we resort to neural representation learning and contrastive learning. Formally, let $x_i^{\mathrm{txt}}$denote the feature vector that represents the accompaniment texture of $x_i^{\mathrm{acc}}$ . It is computed by:

\begin{equation}\label{text_enc}
\begin{aligned}
    x_i^{\mathrm{txt}} = \mathrm{Enc^{txt}_{\theta}}(x_i^{\mathrm{acc}}) ,
\end{aligned}
\end{equation}
where the design of $\mathrm{Enc}_\theta^\mathrm{txt}(\cdot)$  is adopted from the chord-texture representation disentanglement model in \cite{wang2020learning}. This texture encoder regards piano-roll inputs as images and uses a CNN to compute a rough ``sketch'' of polyphonic texture that is not sensitive to mild chord variations. 

To reveal whether two adjacent textures $(x_i^{\mathrm{txt}}, x_{i+1}^{\mathrm{txt}})$ follow a natural transition, we use a contrastive loss $\mathcal{L}$ to simultaneously train the weight matrix $W$ in Eq \eqref{tran} and fine-tune $\mathrm{Enc}_\theta^\mathrm{txt}(\cdot)$ (with parameter $\theta$) in Eq \eqref{text_enc}:
\begin{equation}\label{loss}
    {\mathcal{L}(W, \theta) = 1 - \frac{\exp(\mathrm{sim}(W_1 x_i^{\mathrm{txt}}, W_2 x_{i+1}^{\mathrm{txt}}) )}{\sum_{x \in S} \exp(\mathrm{sim}(W_1 x_i^{\mathrm{txt}}, W_2 x_k^{\mathrm{txt}}) )},}
\end{equation}
where $x_i$ and $x_{i+1}$ are supposed to be consecutive pairs. $S$ is a collection of $k$ samples which contains $x_{i+1}$ and other $k-1$ randomly selected phrases from reference space $r$. Following \cite{bretan2016unit}, we choose $k = 5$.

\begin{figure}
 \centerline{
 \includegraphics[width=\linewidth]{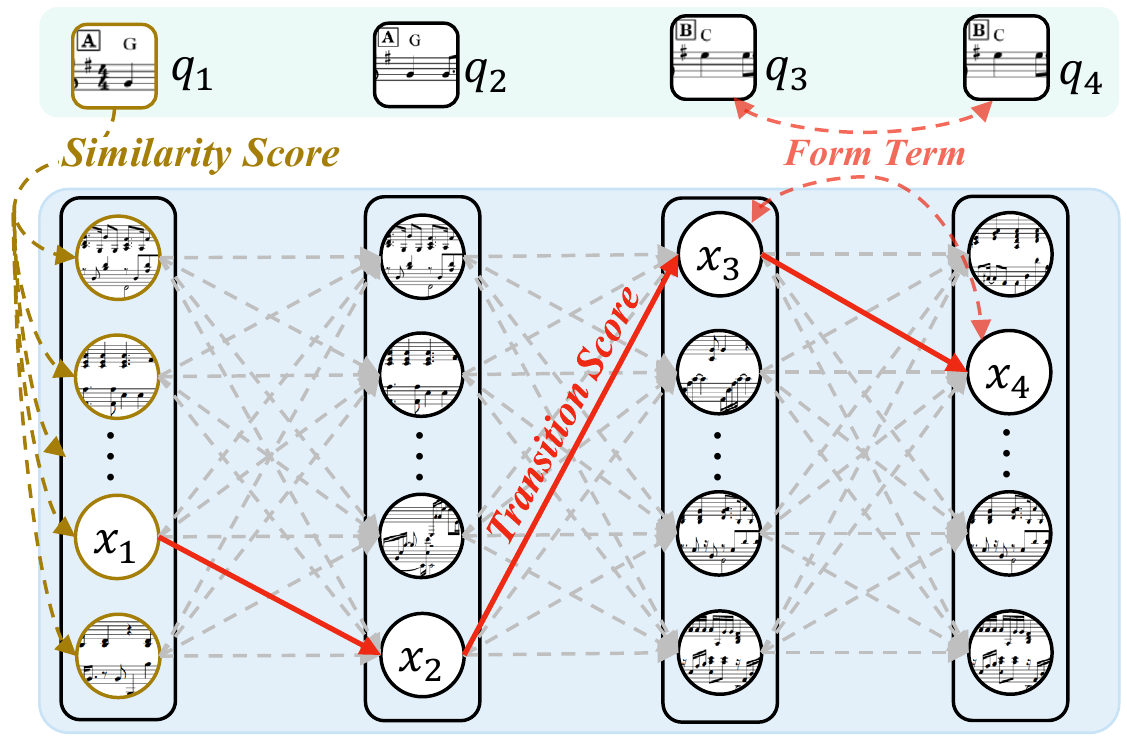}}
 \caption{Phrase selection on the graph. Based on the lead sheet with an {$\mathtt{AABB}$} phrase structure, the search space forms a graph with four consecutive layers. Graph nodes are assigned with similarity scores, and edges with transition scores. The form term is part of the transition score.}
 \label{selection}
\end{figure}

For the form term $\mathrm{form}(x_i, x_{i+1})$, we introduce this term to bias a more well-structured transition. Concretely, if query phrases $q_i$ and $q_{i+1}$ share the same phrase label, we would prefer to also retrieve equal-labeled references, i.e., accompaniments with recapitulated melody themes. To maximize such likelihood, we define the form term:
\begin{equation}
    \mathrm{form}(x_i, x_{i+1}) = \mathbbm{1}_{\{ q_{i}^{\mathrm{label}}=q_{i+1}^{\mathrm{label}}\} } \cdot \mathbbm{1}_{\{x_{i}^{\mathrm{mel}} \approx x_{i+1}^{\mathrm{mel}}\}},
\end{equation}
where we define $x_{i}^{\mathrm{mel}} \approx x_{i+1}^{\mathrm{mel}}$ if and only if their step-wise cosine similarity is greater than 0.99.

\subsubsection{Model Inference}
The reference space forms a layered graph with consecutive layers fully connected to each other. In \figref{selection}, we leverage the transition model to assign weights of edges and the fitness model to assign weights of nodes. Thus, the phrase selection is formulated as:
\begin{equation}\label{optim}
    {\bf x}^* = \underset{x_1, x_2, \cdots, x_{n}}{\mathrm{argmax}} \delta\sum^{n}_{i=1} f(x_i, q_i) + \gamma\sum^{n-1}_{i=1} t(x_i, x_{i+1}),
\end{equation}
where $f(\cdot)$ and $t(\cdot)$ are as defined in Eq \eqref{sim} and Eq \eqref{tran}, and $\delta$ and $\gamma$ are hyper-parameters.

We optimize Eq \eqref{optim} by dynamic programming to retrieve the Viterbi path ${\bf x}^*$ as the optimal solution \cite{forney1973viterbi}. The time complexity is $\mathcal{O}(nN^2)$, where $n$ is the number of query phrases and $N$ is the volume of the reference space.

In summary, the phrase selection algorithm enforces strong structural constraints (song-level form and phrase-level fitness) as well as weak harmonic constraints (chord term in Eq \eqref{sim}) to the selection of accompaniment reference. We argue that this is a good compromise because strong harmonic constraints can potentially ``bury'' well-structured references due to unmatched chord when our dataset is limited. To maintain a better harmonic fitness, we resort to music style transfer.

\subsection{Style Transfer}\label{styletransfer}
The essence of style transfer is to transfer the chord sequence of a selected reference phrase while keeping its texture. To this end, we adopt the chord-texture disentanglement VAE framework by \cite{wang2020learning}. The VAE consists of a chord encoder $\mathrm{Enc}^\mathrm{chd}$ and a texture encoder $\mathrm{Enc}^\mathrm{txt}$. $\mathrm{Enc}^\mathrm{chd}$ takes in a two-bar chord progression under one-beat resolution and exploits a bi-directional GRU to approximate a latent chord representation $z_{\mathrm{chd}}$. $\mathrm{Enc}^\mathrm{txt}$ is introduced in Section \ref{transcore} and it extracts a latent texture representation $z_{\mathrm{txt}}$. The decoder $\mathrm{Dec}$ takes the concatenation of $z_{\mathrm{chd}}$ and $z_{\mathrm{txt}}$ and decodes the music segment using the same architecture invented in \cite{wang2020pianotree}. Sustaining texture input and varying chords, the whole model works like a conditional VAE which re-harmonizes texture based on the chord condition.

In our case, to re-harmonize the selecetd accompaniment $x_i^{\mathrm{acc}}$ to query lead sheet $q_i$, the style transfer works in a pipeline as follows: 
\begin{equation}\label{styletransferequation}
\begin{aligned}
     z_{\mathrm{chd}} &= \mathrm{Enc}^\mathrm{chd}(q_i^{\mathrm{chord}}),\\
     z_{\mathrm{txt}} &= \mathrm{Enc}^\mathrm{txt}(x_i^{\mathrm{acc}}),\\
     x_i^{\prime} &= \mathrm{Dec}(z_{\mathrm{chd}}, z_{\mathrm{txt}}),
\end{aligned}
\end{equation}
where $x_i^{\prime}$ is the re-harmonized result. The final accopmaniment arrangement result is ${\bf x}^{*\prime} = [x^{*\prime}_1, x^{*\prime}_2, \cdots, x^{*\prime}_{n}]$.

\subsection{Controllability}\label{pre-filtering-control}

In the phrase selection stage, we essentially traverse a route on the graph. Intuitively, we can control generation of the whole route by assigning the first node. In our case, we filter reference candidates for the first phrase based on textural properties. The current design has two filter criteria: {\itshape rhythm density} and {\itshape voice number}. we define three intervals {\itshape low}, {\itshape medium}, and {\itshape high} for both properties and mask the references that do not fall in the expected interval. 
\begin{itemize}
    \item Rhythm Density (RD): the ratio of time steps with note onsets to all time steps;
    \item Voice Number (VN): the average number of notes that are simultaneously played.
\end{itemize}

\section{Experiment}

\subsection{Dataset}
We collect our reference space from POP909 dataset \cite{wang2020pop909} with the phrase segmentation created by \cite{dai2020automatic}. POP909 contains piano arrangements of 909 popular songs created by professional musicians, which is a great source of delicate piano textures. Each song has a separated melody, chord, and accompaniment MIDI track. We only keep the pieces with $\frac{2}{4}$ and $\frac{4}{4}$ meters and quantize them at 16th notes (chords at 4th). This derives 857 songs segmented into 11032 phrases. As shown in \tabref{phrase statistics}, we have four-bar and eight-bar phrases in majority, which makes sense for popular songs. We also use POP909 to fine-tune our transition model, during which we randomly split the dataset (at song level) into training (95\%) and validation (5\%) sets.
\begin{table}[t]
  \centering
  \caption{Length Distribution of POP909 Phrase}\label{phrase statistics}
    \begin{tabular}{lcccccc}
    \toprule
    \textbf{bars} & \multicolumn{1}{c}{<4} & 4 & \multicolumn{1}{c}{5\textasciitilde7} & 8     & \multicolumn{1}{c}{>8} \\
    \midrule
    \textbf{Phrases} & 1338   & 3591  & 855   & 3796  & 1402 \\
    \bottomrule
    \end{tabular}%
  \label{tab:addlabel}%
\end{table}%

At inference time, the query lead sheets come from the Nottingham Dataset \cite{nottingham}, a collection of \textasciitilde1000 British and American folk tunes. We also adopt $\frac{2}{4}$ and $\frac{4}{4}$ pieces quantized at 16th (chords at 4th). We label their phrase segmentation by hand, where four-bar and eight-bar phrases are also the most common ones.

\begin{figure*}[t]
 \centerline{
 \includegraphics[width=\linewidth]{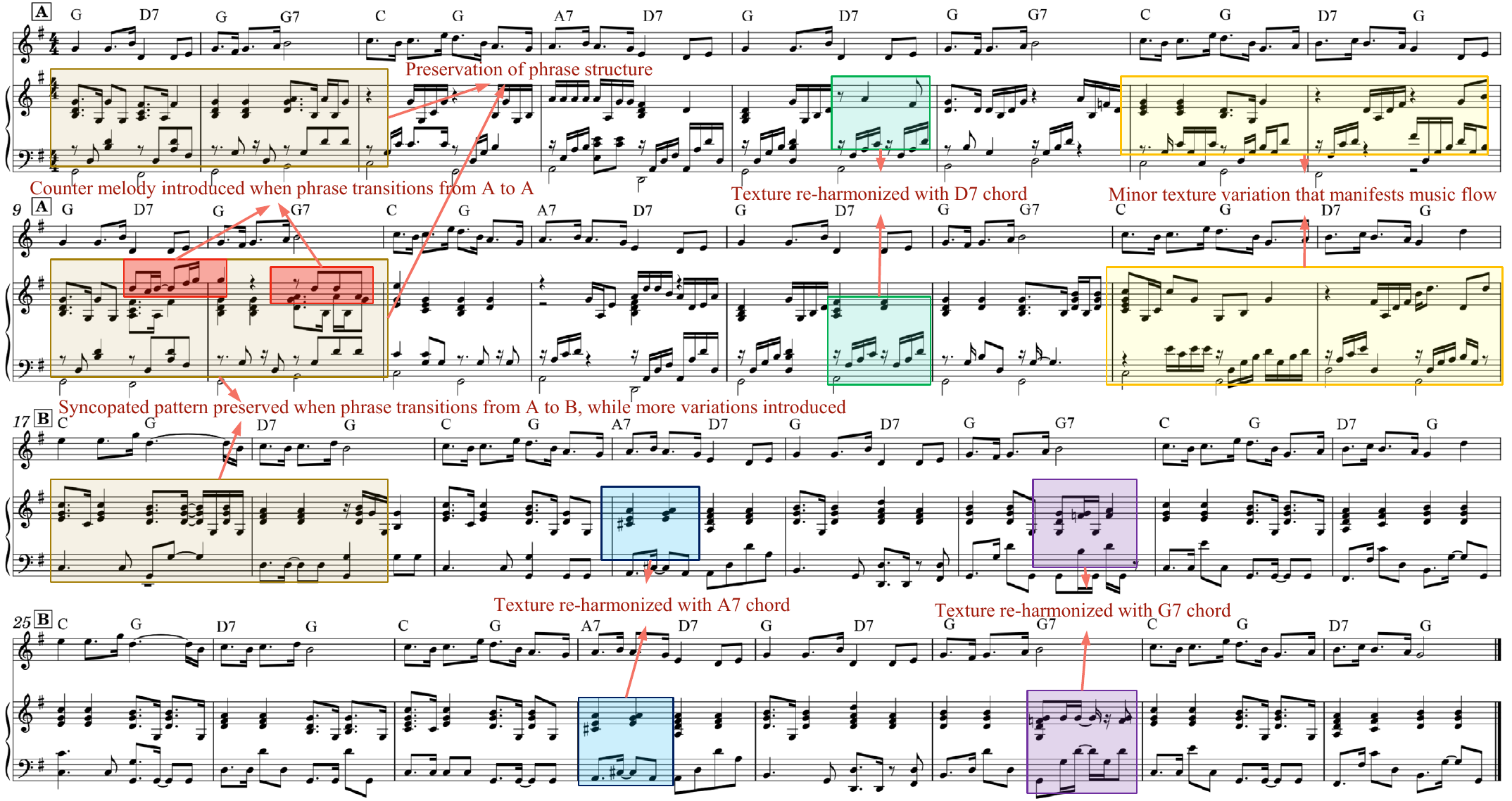}}
 \caption{Accompaniment arrangement for {\itshape Castles in the Air} from Nottingham Dataset by AccoMontage. The 32-bar song has an $\mathtt{A8A8B8B8}$ phrase structure which is captured during accompaniment arrangement. Second melodies and texture variations are also introduced to manifest music flow. Here we highlight some texture re-harmonization of 7th chords.
 }
 \label{sheet}
\end{figure*}
\subsection{Architecture Design}
We develop our model based on the chord-texture disentanglement model proposed by \cite{wang2020learning}, which comprises a texture encoder, a chord encoder, and a decoder. The texture encoder consists of a convolutional layer with kernel size $12\times 4$ and stride $1\times 4$ and a bi-directional GRU encoder \cite{roberts2018hierarchical}. The convolutional layer is followed by a ReLU activation \cite{nair2010rectified} and max-pooling with kernel size $4 \times 1$ and stride $4 \times 1$. The chord encoder is a bi-directional GRU encoder. The decoder is consistent with PianoTree VAE \cite{wang2020pianotree}, a hierarchical architecture for polyphonic representation learning. The architecture of $\mathrm{Enc}^\mathrm{txt}(\cdot)$ in the proposed transition model is the same as the texture encoder illustrated above. We directly take the chord-texture disentanglement model with pre-trained weights as our style transfer model. We fine-tune the transition model with $W_1$ and $W_2$ in Eq \eqref{loss} as trainable parameters.

\subsection{Training}
Our model is trained with a mini-batch of 128 piano-roll pairs for 50 epochs using Adam optimizer \cite{kingma2017adam} with a learning rate from 1e-4 exponentially decayed to 5e-6. Note that each piano-roll pair contains 2 consecutive piano-rolls and 4 randomly sampled ones. We first pre-train a chord-texture disentanglement model and initialize $\mathrm{Enc}^\mathrm{txt}(\cdot)$ using weights of the texture encoder in the pre-trained model. Then we update all the parameters of the proposed transition model using contrastive loss $\mathcal{L}$ in Eq \eqref{loss}. We set both $\alpha$ and $\beta$ in Eq \eqref{sim} to 0.5. 
During inference, we set $\delta$ and $\gamma$ in Eq \eqref{optim} to $0.3$ and $0.7$.  

\begin{figure*}
 \centerline{
 \includegraphics[width=2.0\columnwidth]{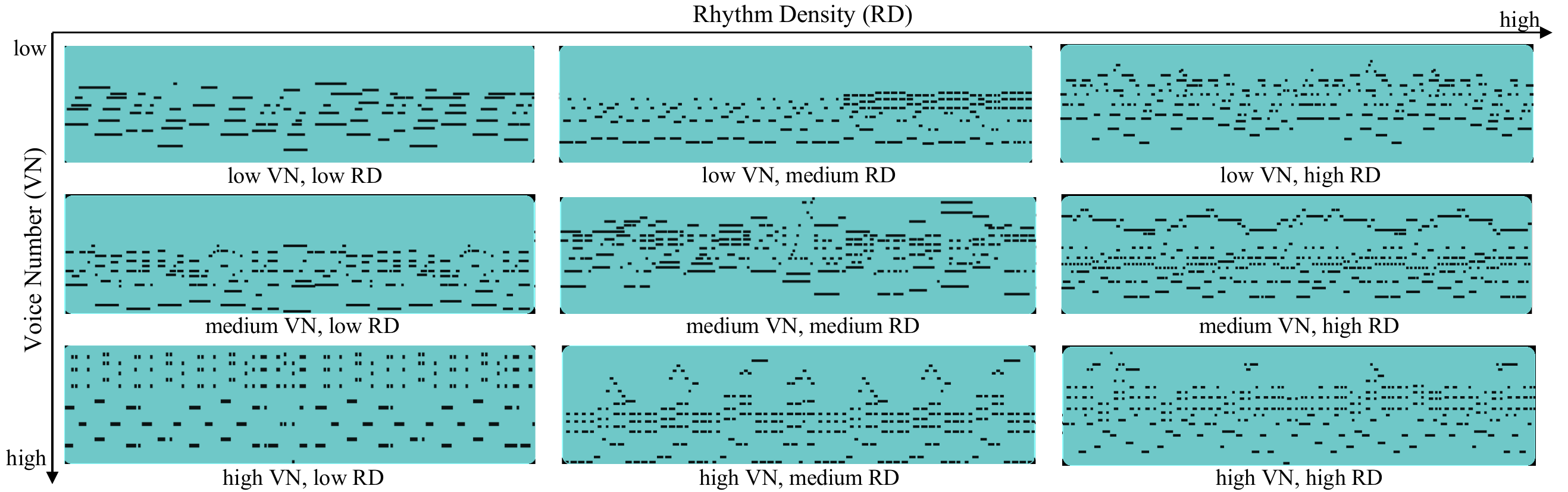}}
 \caption{Pre-Filtering Control on Rhythm Density and Voice Number}
 \label{control}
\end{figure*}
\subsection{Generated Examples}
To this end, we show two long-term accompaniment arrangement examples by the Accomontage system. The first one is illustrated in Figure~\ref{sheet}, in which we show a whole piece (32-bar music) piano arrangement (the bottom two staves) base on the lead sheet (the top stave). We see that the generated accompaniment matches with the melody and has a natural flow on its texture. Moreover, it follows the {$\mathtt{A8A8B8B8}$} structure of the melody.

The second example shows that our controls on rhythm density and voice number are quite successful. To better illustrate, we switch to a piano-roll representation in Figure~\ref{control}, where 9 arranged accompaniments for the same lead sheet is shown in a 3 by 3 grid. The rhythm density control increases from left to right, while the voice number control increases from top to bottom. We can see that both controls have a significant influence on the generated results.

\subsection{Evaluation}
\subsubsection{Baseline Models}
The AccoMontage system is a generalized template-based model that leverages both rule-based optimization and deep learning to complement each other. To evaluate, we introduce a hard template-based and a pure learning-based baseline to compare with our model. Specifically, the baseline model architectures are as follows:

\textbf{Hard Template-Based (HTB)}: The hard template-based model also retrieves references from existing accompaniment, but directly applies them without any style transfer. It uses the same phrase selection architecture as our model while skipping the style transfer stage.

\textbf{Pure Learning-Based (PLB)}: We adopt the accompaniment arrangement model in \cite{wang2020learning}, a seq2seq framework combining Transformer \cite{vaswani2017attention} and chord-texture disentanglement. We consider  \cite{wang2020learning} the current state-of-the-art algorithm for controllable accompaniment generation due to its tailored design of harmony and texture representations, sophisticated neural structure, and convincing demos. The input to the model is a lead sheet and its first four-bar accompaniment. The model composes the rest by predicting every four bars based on the current lead sheet and previous four-bar accompaniment.

\subsubsection{Subjective Evaluation}
We conduct a survey to evaluate the musical quality of the arrangement performance of all models. In our survey, each subject listens to 1 to 3 songs randomly selected from a pool of 14. All 14 songs are randomly selected from the Nottingham Dataset, 12 of which have 32 bars and the other two 24 and 16 bars. Each song has three accompaniment versions generated by our and the baseline models. The subjects are required to rate all three accompaniment versions of one song based on three metrics: coherence, structure, and musicality. The rating is base on a 5-point scale from 1 (very poor) to 5 (excellent).
\begin{itemize}
    \item \textbf{Coherence}: If the accompaniment matches the lead melody in harmony and texture;
    \item \textbf{Structure}:If the accompaniment flows dynamically with the structure of the melody;
    \item \textbf{Musicality}: Overall musicality of accompaniment.
\end{itemize}
\begin{figure}
 \centerline{
 \includegraphics[width=0.9\columnwidth]{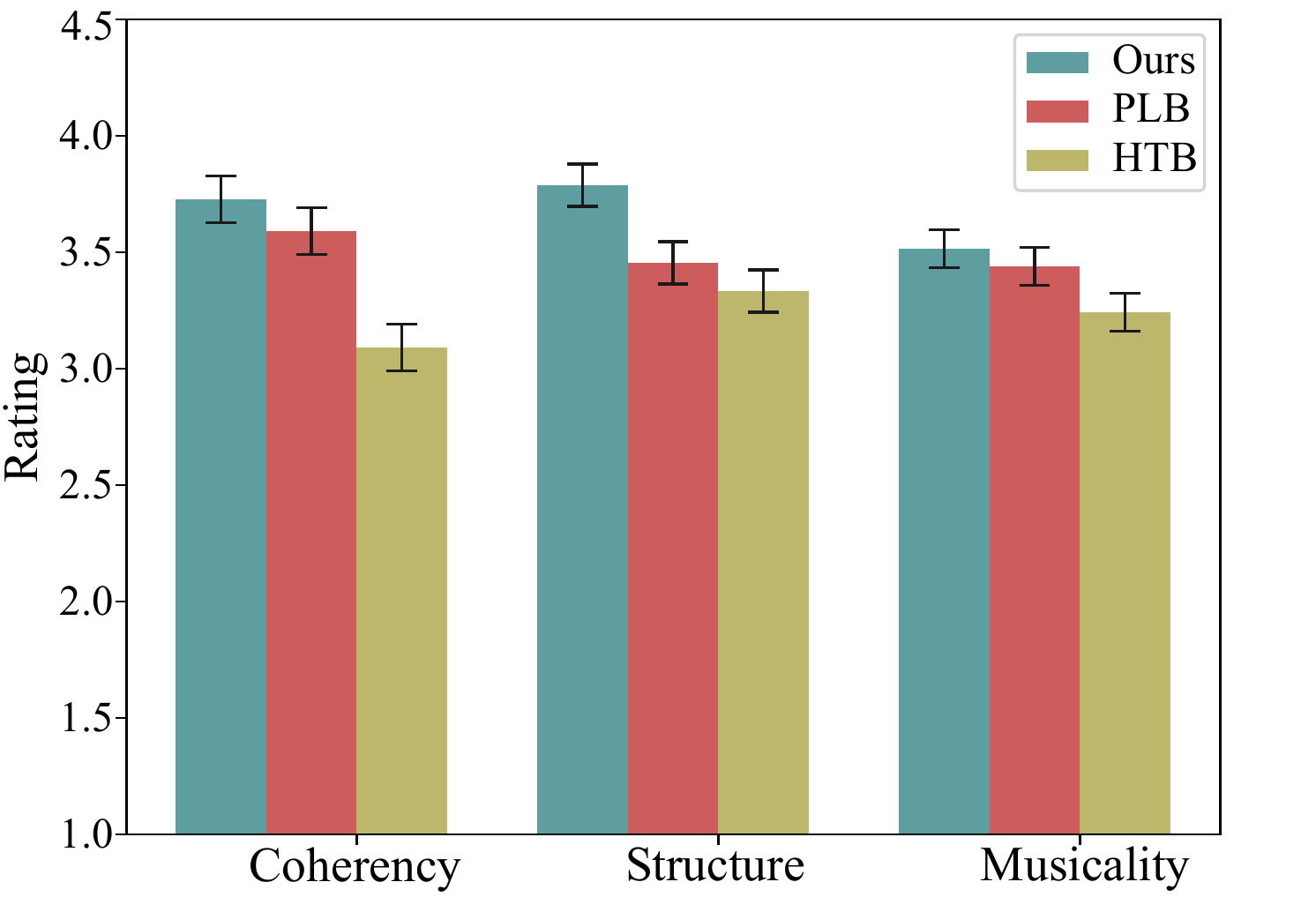}}
 \caption{Subjective Evaluation Results.}
 \label{subject}
\end{figure}

A total of 72 subjects (37 females and 35 males) participated in our survey and we obtain 67 effective ratings for each metric. As in \figref{subject}, the heights of bars represent the means of the ratings and the error bars represent the MSEs computed via within-subject ANOVA \cite{scheffe1999analysis}. We report a significantly better performance of our model than both baselines in coherence and structure ($p<0.05$), and a marginally better performance in musicality ($p=0.053$).

\subsubsection{Objective Evaluation}
In the phrase selection stage, we leverage a self-supervised contrastive loss (Eq \eqref{loss}) to enforce a smooth textural transition among reference phrases. We expect a lower loss for true adjacent phrase pairs than in other situations. Meanwhile, true consecutive pairs should hold a similar texture pattern with smaller differences in general properties. 

\begin{figure}[t]
 \centerline{
 \includegraphics[width=0.9\columnwidth]{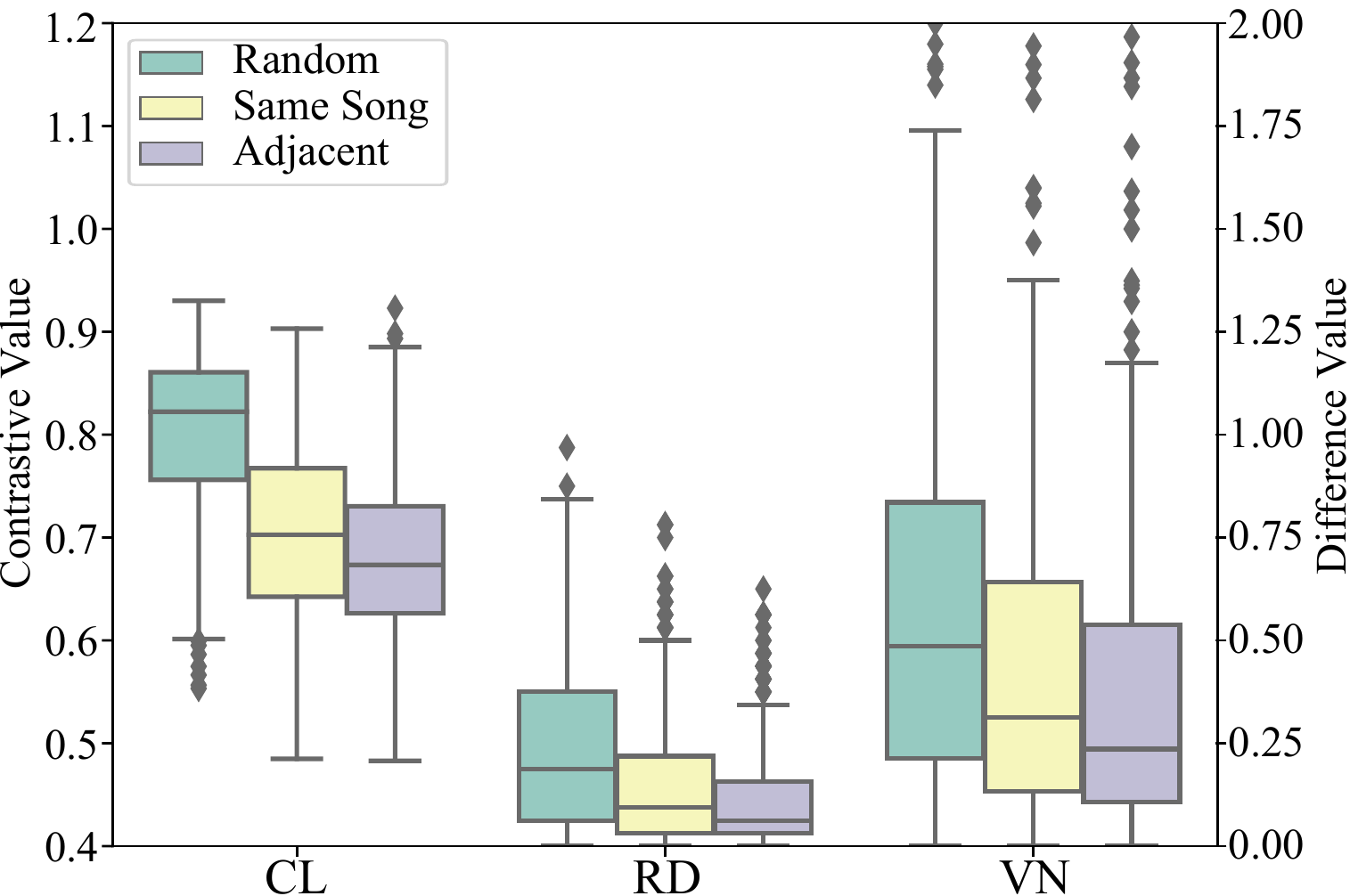}}
 \caption{Evaluation of Transition Model. The contrastive loss (CL) and differences of RD and VN are calculated for three types of phrase pairs. A consistent decreasing trend illustrates reliable
 discernment of smooth transition.}
 \label{cl}
\end{figure}

We investigate the contrastive loss (CL) and the difference of rhythm density (RD) and voice number (VN) among three types of phrase pairs from the validation set. Namely, {\itshape Random}, {\itshape Same Song}, and {\itshape Adjacent}. Between totally randomly pairing and strict adjacency, {\itshape Same Song} refers to randomly selecting two phrases (not necessarily adjacent) from one song. Results are shown in Figure~\ref{cl}.

\begin{table}[t]
  \centering
  \caption{Ranking Accuracy and Mean Rank}
    \begin{tabular}{cccc}
    \toprule
    \textbf{Metric} & Phrase Acc. & Song Acc. & \multicolumn{1}{l}{Rank@50} \\
    \midrule
    \textbf{Value} & 0.2425 & 0.3769 & 5.8003 \\
    \bottomrule
    \end{tabular}%
  \label{mrank}%
\end{table}%

For contrastive loss and each property, we see a consistent decreasing trend from {\itshape Random} to {\itshape Same Song} and to {\itshape Adjacent}. Specifically, we see the upper quartile of {\itshape Adjacent} is remarkably lower than the lower quartile of {\itshape Random} for CL, which indicates a reliable textural discernment that ensures smooth phrase transitions. This is also proved by the metric of ranking accuracy and mean rank \cite{bretan2016unit}, where the selection rank of the true adjacent phrase out of $k-1$ randomly selected phrases (Rank@$k$) is calculated. We follow \cite{bretan2016unit} and adopt Rank@$50$, and the results are shown in \tabref{mrank}. Phrase Acc. and Song Acc. each refers to the accuracy that the top-ranked phrase is {\itshape Adjacent} or belongs to the  {\itshape Same Song}. The high rank of adjacent pairs illustrates our model's reliability to explore smooth transitions.

\section{Conclusion}
In conclusion, we contribute a generalized template-based algorithm for the accompaniment arrangement problem. The main novelty lies in the methodology that seamlessly combines deep generation and search-based generation. In specific, searching is used to optimize the high-level structure, while neural style transfer is in charge of local coherency and melody-accompaniment fitness. Such a top-down hybrid strategy is inspired by how human musicians arrange accompaniments in practice. We aim to bring a new perspective not only to music generation, but to long-term sequence generation in general. Experiments show that our AccoMontage system significantly outperforms pure learning-based and template-based methods, being capable of rendering well-structured and fine-grained accompaniment for full-length songs. 

\section{Acknowledgement}
The authors wish to thank Yixiao Zhang for his contribution to figure framing and proofreading. We thank Liwei Lin and Junyan Jiang for providing feedback on initial drafts of this paper and additional editing.

\bibliography{AccoMontage.bib}

%
%
%
%
%

\end{document}